\documentclass{emulateapj}

\shorttitle{Cosmic star formation history based on radio data}
\shortauthors{V. Smol\v{c}i\'{c} et al.}

\def\f#1   {Fig.~\ref{#1}}
\def\s#1   {Sec.~\ref{#1}}
\def\tab#1   {Tab.~\ref{#1}}
\def\t#1   {Tab.~\ref{#1}}
\def\lum   {$\mathrm{L}_{1.4}$}
\def\comm#1   {{\tt (COMMENT: #1) }}

\def\sqdeg            {$\Box^{\circ}$}
\def\sqdegs           {$\Box^{\circ}$}

\def\lsun              {$\mathrm{L}_{\odot}$}
\def\Msol              {$\mathrm{M}_{\odot}$}

\def\Msolyr              {$\mathrm{M}_{\odot}\, \mathrm{yr}^{-1}$}
\def\WH                {W~Hz$^{-1}$}
\def\wh                {W~Hz$^{-1}$}
\def\mic               {$\mathrm{\mu m}$}

\def\Vm                {$V_\mathrm{max}$}

\slugcomment{  }

\begin{document}

\title{ The dust un-biased cosmic star formation history from the 20~cm
  VLA-COSMOS survey\altaffilmark{0}}

\author{V.~Smol\v{c}i\'{c}\altaffilmark{1,2,3},
        E.~Schinnerer\altaffilmark{1}, 
        G.~Zamorani\altaffilmark{4},
        E.~F.~Bell\altaffilmark{1},
        M.~Bondi\altaffilmark{5}
        C.~L.~Carilli\altaffilmark{6},
        P.~Ciliegi\altaffilmark{4}
        B.~Mobasher\altaffilmark{7}
        T.~Paglione\altaffilmark{8,9},
        M.~Scodeggio\altaffilmark{10},
        N.~Scoville\altaffilmark{3}
        }
\altaffiltext{0}{Based on observations with the National
Radio Astronomy Observatory which is a facility of the National Science
Foundation operated under cooperative agreement by Associated Universities,
Inc. }
\altaffiltext{1}{Max Planck Institut f\"ur Astronomie, K\"onigstuhl 17,
  Heidelberg, D-69117, Germany } 
\altaffiltext{2}{Fellow of the International Max Planck Research School for
  Astronomy and Cosmic Physics}
\altaffiltext{3}{California Institute of Technology, MC 105-24, 1200 East
California Boulevard, Pasadena, CA 91125 }
\altaffiltext{4}{INAF - Osservatorio Astronomico di Bologna, via Ranzani 1,
  40127, Bologna, Italy} 
\altaffiltext{5}{INAF - Istituto di Radioastronomia, via Gobetti 101, 40129
  Bologna, Italy }
\altaffiltext{6}{National Radio Astronomy Observatory, P.O. Box 0, Socorro,
  NM 87801-0387 } 
\altaffiltext{7}{Physics and Astronomy Department, University of California, Riverside, 
900 University Ave, CA 92521, USA}
\altaffiltext{8}{York College, City University of New York, 94-20 Guy R. Brewer
  Boulevard, Jamaica, NY 11451} 
\altaffiltext{9}{American Museum of Natural History,
Central Park West at 79th Street, New York, NY 10024} 
\altaffiltext{10}{IASF Milano-INAF, Via Bassini 15, I-20133, Milan, Italy }

\begin{abstract}
 We derive the cosmic star formation history (CSFH) out to $z = 1.3$
 using a sample of $\sim350$ radio-selected star-forming galaxies, a
 far larger sample than in previous, similar studies.  We attempt to
 differentiate between radio emission from AGN and star-forming
 galaxies, and determine an evolving 1.4\,GHz luminosity function
 based on these VLA-COSMOS star forming galaxies.  We precisely
 measure the high-luminosity end of the star forming galaxy luminosity
 function (SFR $\ga 100\,M_{\sun}\,{\rm yr}^{-1}$; equivalent to
 ULIRGs) out to $z = 1.3$, finding a somewhat slower evolution than
 previously derived from mid-infrared data.  We find that more stars
 are forming in luminous starbursts at high redshift. We use
 extrapolations based on the local radio galaxy luminosity function;
 assuming pure luminosity evolution, we derive $L_* \propto (1+z)^{2.1
 \pm 0.2}$ or $L_* \propto (1+z)^{2.5 \pm 0.1}$, depending on the
 choice of the local radio galaxy luminosity function.  Thus, our
 radio-derived results independently confirm the $\sim 1$ order of
 magnitude decline in the CSFH since $z \sim 1$.
\end{abstract}


\section{Introduction}

Studies based on different galaxy star formation indicators (UV,
optical, FIR, radio) agree that the cosmic star formation history
(i.e.\ the total star formation rate per unit co-moving volume; CSFH
hereafter) has declined by about an order of magnitude since $z\sim1$ (for a
compilation see e.g.\ \citealt{hopkins04}). One of the major
difficulties of UV/optical based tracers is the significant
model-dependent dust-obscuration correction that needs to be imposed
on the data. This `dust-obscuration problem' may be overcome using
longer wavelengths, such as the IR and radio regimes.  However, in
these cases a multi-wavelength approach is essential as redshift
information and a reliable identifier of star forming (SF) galaxies is
required (e.g.\ \citealt{caputi07, smo08}; S08 hereafter).  In this
context the {\em radio} star formation tracer provides an important
complementary view of the CSFH. First, radio emission is a
dust-insensitive tracer of recent star formation (not affected
by old stellar populations; see \citealt{condon92} for a review).
Second, interferometric radio observations with $\sim1''$ resolution
allow more reliable identifications (compared to FIR and sub-mm data)
with objects detected at other wavelengths.

The dust-unbiased total CSFH has been constrained recently using MIR
(24/8\mic) selected samples obtained by deep small area surveys (CDFS,
GEMS, GOODS; \citealt{lefloc'h05, zheng06, zheng07, caputi07,
bell07}). Small area surveys, however, are subject to cosmic
variance. Moreover, they do not observe a large enough comoving volume
in order to fairly sample rare high-luminosity galaxies.   In this
paper we use the 2\sqdeg\ COSMOS field \citep{scoville07a}, and its
1.4~GHz radio observations \citep{schinnerer07}, to derive the cosmic
star formation history. In such a large field cosmic variance is
significantly reduced as 2\sqdegs\ ($1.4^\circ\times1.4^\circ$) sample
comoving volumes in the early universe ($z\sim1$) comparable to the
largest survey in the local universe (SDSS -- DR1; see Fig.~1 in
\citealt{scoville07a}). The physical angular size sampled by
1.4$^\circ$ at redshifts 0.2 -- 1.1 roughly corresponds to a factor of
3 to 8 of the typical cluster scale length ($\sim5$~Mpc). Thus at all
redshifts, such a field fairly samples relevant structures in the
universe (see \citealt{scoville07a, scoville07b, mccracken07} for a
more detailed discussions on cosmic variance in the COSMOS field).

In the last decade several radio surveys have been utilized to
independently derive the cosmic star formation history. However, their
results are based on  small observed areas, non-uniform rms in the final map,
as well as a fairly non-uniform selection of SF galaxies.  The first
derivation of the CSFH based on radio data has been performed by
\citet{haarsma00}. They combined three radio frequency observations of
the Hubble Deep Field ($66$~arcmin$^2$; \citealt{richards98}), SSA13
($7$~arcmin$^2$; \citealt{windhorst95}), and V15 ($86$~arcmin$^2$;
\citealt{fomalont91, hammer95}) fields reaching $5\sigma$
radio depths of 9, 8.8, and 16~$\mu$Jy at the field centers,
respectively. Of the total number of their radio-selected sources (77)
only 37 were securely classified as star forming galaxies (based on
morphology and/or optical spectroscopy). Their sources reach out to
$z\sim3$, 23 have spectroscopic redshifts, and the redshifts for the
remaining 14 have been estimated using only I or H and K' bands (see
\citealt{haarsma00} for details). More recently, \citet{seymour08}
used radio observations of the 13$^\mathrm{H}$ XMM-Newton/Chandra Deep
field (0.196\sqdeg ; $4\sigma\sim30~\mu$Jy ) to derive the CSFH. 
Out of a total of 449 radio sources they
find 269 galaxies which they classify as star
forming based on a number of criteria applicable to sub-samples of
their objects (see their Tab.~1). About half of these galaxies have a
spectroscopic redshift ($z\lesssim3$).

Here we utilize the 1.4~GHz VLA observations of the COSMOS 2\sqdeg\
field \citep[VLA-COSMOS Large Project;][]{schinnerer07} to overcome the
above mentioned biases. The final mosaic has a resolution of $1.5''
\times 1.4''$ and a typical $rms$ of $\sim 10.5~(15)$~$\mu$Jy/beam in
the central 1 (2) \sqdegs\ making this survey to date the largest
radio deep field at this sensitivity and angular resolution. Given the
COSMOS panchromatic data set, \citet{smo08} have developed a novel
method to select star forming and AGN galaxies based only on NUV--NIR
photometry. This yielded a robust statistical selection of 340 SF
galaxies out to $z=1.3$ (see \s{sec:sample} ). One particular
advantage of the VLA-COSMOS survey are the large comoving volumes
observed at all redshifts, thus allowing one to study a statistically
significant sample of rare objects, such as the most intensely star
forming galaxies.  In this paper we focus on the derivation of the
CSFH, with emphasis on the evolution of galaxies with high star
formation rates ($\gtrsim100$~\Msol~yr$^{-1}$), using the 1.4~GHz
VLA-COSMOS Large Project. 


We report magnitudes in the AB system, adopt a standard concordance
cosmology ($H_0=70,\, \Omega_M=0.3, \Omega_\Lambda = 0.7$), and define
the radio synchrotron spectrum as $F_{\nu} \varpropto \nu^{-\alpha}$,
assuming $\alpha = 0.7$.

\section{ The 1.4~GHz luminosity function for star forming galaxies in
  VLA-COSMOS} 
\label{sec:lf}

\subsection{ The star forming galaxy sample } 
\label{sec:sample}

The sample of SF galaxies used here is presented in S08, and briefly
summarized below.  Using radio and optical data for the COSMOS field
S08 have constructed a sample of 340 star forming galaxies with
$z\leq1.3$ out of the entire VLA-COSMOS catalog. The selection required
optical counterparts with $i\leq26$, accurate photometry (i.e.\
outside photometrically flagged areas), and a $\mathrm{S/N}\geq5$
detection at 20~cm, and is based on rest-frame optical colors (see
also \citealt{smo06}).  The classification method was well-calibrated
using a large local sample of galaxies ($\sim7,000$ SDSS ``main''
spectroscopic galaxy sample, NVSS and IRAS surveys) representative of
the VLA-COSMOS population. It was shown that the method agrees
remarkably well with other independent classification schemes based on
mid-infrared colors and optical spectroscopy, and that, within the
selection limits, it is not biased against dusty starburst
galaxies. It is important to note that the rest-frame colors for the
VLA-COSMOS radio sources have been shown to stay basically unchanged
with redshift (an effect that was also noted by
\citealt{barger07,huynh08}).

Out of the 340 selected SF galaxies 150 have spectroscopic redshifts
available, while the remaining sources have very reliable photometric
redshifts (see S08 and references therein).  Based on Monte Carlo
simulations, S08 have shown that the photometric errors in the
rest-frame color introduce a small, $\sim5\%$, number incompleteness
of SF galaxies.  Here we use the S08 sample of SF galaxies,
statistically corrected for this effect.

\subsection{Derivation of the radio luminosity function (LF)}

We derive the radio luminosity function ($\Phi$) for our SF galaxies
in four redshift bins
using the standard $1/$\Vm\ method \citep{schmidt68}. 
In computing $V_\mathrm{max}$ we take into account both the radio and
optical flux limits (i.e.\ the maximum observable redshift, $z_\mathrm{max}$),
as well as the non-uniform rms noise level in the VLA-COSMOS mosaic. For the
latter we use the differential visibility area (i.e.\ areal coverage, $A_k$,
vs.\ rms; see Fig.~13 in \citealt{schinnerer07}). For a source with 1.4~GHz
luminosity $L_j$ its maximum volume is then $V_\mathrm{max}(L_j) =
\sum_{k=1}^{n} A_k \cdot V_\mathrm{max}(z_\mathrm{max}^{A_k},L_j)$.

Several additional corrections need to be taken into account: 
i) the VLA-COSMOS detection completeness, ii) the fraction of sources not
included in the radio-optical sample, and iii) the SF galaxy selection bias
due to the rest-frame color uncertainties.

 The radio detection completeness of the VLA-COSMOS survey has been
derived by \citet{bondi08} via Monte Carlo -- artificial source --
simulations for the inner 1\sqdegs.  Artificial radio sources have
been simulated taking into account both the flux density and angular
size distributions. The first has been described using a typical
broken power law, while the latter has been assumed to depend on flux
density ($\Theta_\mathrm{med}\propto F_\nu^m$)\footnote{The angular
size dependence of radio sources on their flux density has been
observed in numerous radio surveys, and it is mainly due to the radio
luminosity -- size relation. At fainter flux densities, intrinsically
fainter (and therefore smaller) radio sources are preferentially
sampled. } Comparisons between the median angular sizes in different
flux ranges and between real and simulated 1.4~GHz radio data showed
that the median angular size ($\Theta_\mathrm{med}$) of radio sources
is changing with flux according to a power law with an exponent of
$m=0.5$.  $\Theta_\mathrm{med}\propto F_\nu^{0.5}$ yields flux
dependent completeness corrections. The median of these corrections is
$\sim10\%$ (reaching a maximum value of 60\% in only one of the lowest
flux density bins; see Tab.~1 in \citealt{bondi08}). Assuming that
their corrections, scaled for the differences in radio sensitivity,
hold for the full 2\sqdegs, we utilize them to correct our LFs for the
detection incompleteness ($\mathrm{f_{det}}$).

 To correct for objects located in photometrically-flagged regions in
  the optical images, or not identified with an optical counterpart,
  and thus not included in the radio-optical sample, we construct a
  correction curve as a function of total 1.4~GHz flux density which
  yields an average correction of $\sim30\%$ (see Fig.~23 in S08;
  $\mathrm{f_{flag}}$).  Hence, in the i$^\mathrm{th}$ luminosity bin
  the comoving space density ($\Phi_i$), and its corresponding error
  ($\sigma_i$), are computed by weighting the contribution of each
  galaxy by the two correction factors, $\mathrm{f_{det}}$ and
  $\mathrm{f_{flag}}$, which were obtained by linearly interpolating
  the two correction curves described above, respectively, at the
  total flux of the given, j$^\mathrm{th}$, source:
\begin{eqnarray} 
\Phi_i = \sum_{j=1}^{N} 
              \frac{\mathrm{f_{det}^j}\cdot\mathrm{f_{flag}^j}  }
                   {\mathrm{V_{max}^j}} \,\, ; \quad
\sigma_i = \sqrt{ \sum_{j=1}^{N} 
              \left( 
               \frac{\mathrm{f_{det}^j}\cdot\mathrm{f_{flag}^j}  }
                   {\mathrm{V_{max}^j}} 
               \right)^2
              }
\end{eqnarray}

The selection bias due to the rest-frame color uncertainties is
accounted for via Monte Carlo simulations.  As described in S08, the
rest-frame color error distribution is simulated using a Gaussian
function with $\sigma=0.1$. These errors are then added to the
rest-frame color obtained from the SED fitting. The SF galaxies are
then selected, and the LF is derived as described above.  This process
is repeated 300 times, hence we obtain 300 realizations of
($\Phi_i$,$\sigma_i$) for each luminosity bin. We take the median
values as representative.  

\subsection{The radio luminosity function }

The LFs for our SF galaxies for the $4$ chosen redshift bins are shown in
\f{fig:lf} , and tabulated in \t{tab:LFs} .
In each panel in \f{fig:lf} \ we show the analytical form of the
locally derived 20~cm LF for SF galaxies given by \citet[see also
  \s{sec:evolv} ]{sadler02}.
It is worth noting that although the 2\sqdeg\ COSMOS field samples a
relatively small comoving volume at the lowest redshifts and only
a photometric identification of SF galaxies has been used, our LF in the
lowest redshift bin agrees remarkably well with the local LFs that
were derived using all-sky radio surveys (NVSS) combined with good quality
optical spectroscopic data (SDSS, 2dF) to identify SF galaxies. For
comparison, in the first redshift bin we also indicate other local
radio LFs that exist in the literature \citep{condon89, best05}. We discuss
the implications of the differences between the various local
radio LFs on the star formation history further below (\s{sec:locLFs} ).

\begin{figure}[ht!]
\center{
\includegraphics[bb = 80 330 500 782, width=\columnwidth]{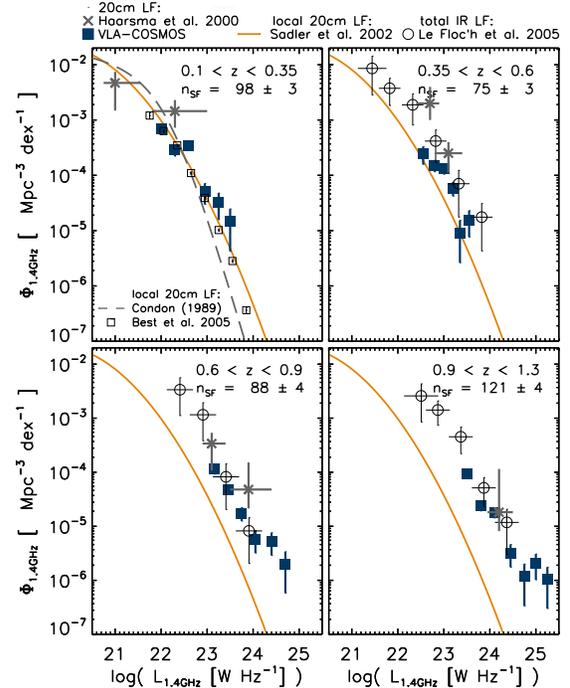}
\caption{ 1.4~GHz luminosity functions (LFs) for star forming galaxies
  in the VLA-COSMOS survey, shown for four redshift ranges (filled
  blue squares) are presented in each panel. The number of galaxies in
  each redshift bin, statistically corrected for selection
  uncertainties (see text for details), is also indicated in each
  panel. The local 20~cm LFs for star forming galaxies are shown
  in the top-left panel (\citealt{condon89, sadler02, best05}; the
  \citealt{sadler02} LF -- solid orange curve -- is shown in all
  panels).   In all panels we show the volume densities derived by
  \citet[gray asterisk]{haarsma00}, corrected to the current
  cosmology.  In the top right, and bottom panels, the total IR LFs
  \citep{lefloc'h05} for the corresponding redshift ranges (open
  circles) is also shown (see text for details). The total IR
  luminosity was converted to 1.4~GHz luminosity using the correlation
  given in \citet{bell03}.
  \label{fig:lf}}
}
\end{figure}

In all panels we show the volume densities of 20~cm radio sources
  derived by \citet[gray asterisk]{haarsma00}, corrected to the current
  cosmology. Haarsma et al.\ used 37 star forming galaxies to derive
  these LFs (38\% of these had approximate redshifts derived from I- or HK'-
  band magnitudes, the others had spectroscopic redshifts). Their data
  points in each redshift range agree fairly well within the
  error-bars with our results. Note that due to the almost one order
  of magnitude larger sample of star forming galaxies used here, the
  error-bars of the VLA-COSMOS LFs are significantly smaller.

In the higher redshift panels we also compare our LFs with the total
IR LFs derived by \citet[hereafter LF05]{lefloc'h05} based on a
24~\mic\ selected sample in the CDFS (Chandra Deep Field South; top
right and both bottom panels in \f{fig:lf} ).  The total IR luminosity
was converted to 1.4~GHz luminosity using the total IR -- radio
correlation \citep{bell03}, which has an intrinsic scatter of
$\sim0.26$~dex, shown by horizontal error bars in \f{fig:lf} . The IR
LFs were re-scaled to our redshift ranges either by combining two
narrower redshift bins given in LF05 or by scaling a given comoving
density using the evolution parameters, and their corresponding
errors, given in LF05.  
There is an excellent agreement between the 1.4~GHz and IR LFs. Note
also that the VLA-COSMOS LFs constrain much better than the IR data
the high-luminosity end, i.e.\ the most intensely star-forming
galaxies.  Interestingly, the VLA-COSMOS star forming galaxies in our
two highest redshift bins ($z>0.6$) show an extended high-luminosity
(\lum~$\gtrsim2\times10^{24}$~\wh ) tail.  We cannot exclude some
possible contamination from AGN, which are much more numerous than SF
galaxies at these high radio luminosities (see Fig.~17 in S08).
However, a similar excess of SF galaxy volume densities at the high
luminosity end has been recently found by Cowie et al. (2004), who
have used the HDF-N and SSA~13 fields to derive the radio LF for {\em
spectroscopically} identified SF galaxies (106 SF galaxies,
$z\leq1.4$).

\subsection{ Towards the derivation of the cosmic star formation history }
\label{sec:evolv}

The comparison of our derived LFs (see \f{fig:lf} ) with the local
20~cm LF shows a strong evolution with look-back time.  The evolution
of a given population of objects is usually parameterized with
monotonic density and luminosity evolution.  However, as the
VLA-COSMOS data do not allow the derivation of the LF out to, and
fainter than the characteristic luminosity (i.e.\ the 'knee' of the
LF), a full determination of the evolution is not possible with these
data, and we choose to parameterize it using pure luminosity
evolution. However, as we show below, its quantization depends fairly
strongly on the choice of the local LF, as those presented in the
literature are not exactly the same.

\subsubsection{The local 20~cm luminosity functions}
\label{sec:locLFs}

The two commonly used analytical forms for the local 20~cm radio LFs
have been presented in \citet{condon89} and \citet{sadler02}. Condon
et al.\ use a hyperbolic parameterization (see also \citealt{condon02})
of the form:

\begin{eqnarray}
\label{eq:condon}
\log \Phi_\mathrm{1.4GHz} & = &  28.83 - \frac{3}{2} \log L_\mathrm{1.4GHz}  + Y \nonumber \\
     & & - \left[ B^2 +\frac{\left( \log L_\mathrm{1.4GHz} - X \right) ^2}{W^2}  \right]^{1/2} 
\end{eqnarray}

where $Y=2.88$, $B=1.5$, $X=22.108$, $W=0.667$ (corrected to the
current cosmology and the base of $d \log L$, compared to $d\,
\mathrm{mag}$ given in \citealt{condon89}). These parameters have been
derived using a sample of 307 spiral and irregular galaxies from the
Revised Shapley-Ames Catalog \citep[RSA;][]{sandage81} observed at
1.49~GHz \citep[see \f{fig:lfcomp} ]{condon87}.

\begin{figure}[h!]
\includegraphics[bb =  54 360 486 792, width=\columnwidth ]{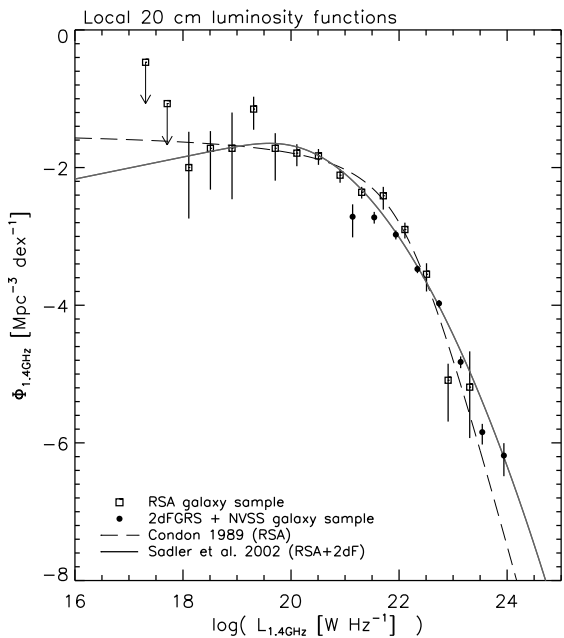}
\caption{ Illustration of the differences in the shapes of the local
    1.4~GHz radio luminosity functions (LFs) found in the literature
    \citep[corrected to the current cosmology]{condon89,
    sadler02}. Also shown are the data points used for the analytical
    fits in \citet[Revised-Shapley-Ames Catalogue; RSA]{condon89} and
    \citet[NVSS/2dFGRS and RSA samples]{sadler02}. Note that the major
    difference in the representation of the local radio LFs is the
    analytical form used to fit the data (see
    eqs.~\ref{eq:condon}~\&~\ref{eq:sadler}) that particularly affects
    the position of the LF break (i.e.\ 'knee').
  \label{fig:lfcomp}}
\end{figure}

 The LF given in \citet{sadler02} takes on the form of a combined
power-law and Gaussian distribution given by the following
analytic function (first proposed by \citealt{sandage79}):
\begin{equation}
\label{eq:sadler}
\Phi(L) = \Phi^* \left[ \frac{L}{L_*} \right]^{1-\alpha}
   \exp{ \left\{ -\frac{1}{2\sigma^2} \left[ \log{ ( 1+\frac{L} {L_*} )}
   \right]^2 \right\} } 
\end{equation}
with $\alpha = 0.84$, $\sigma=0.94$, $\Phi^* =
22.9\times10^{-3}$~Mpc$^{-3}$, and $L^*=1.95\times10^{19}$~W~Hz$^{-1}$
(scaled to the cosmology used here, and to the base of $d \log L$; see
\citealt{hopkins04}).  \citet{sadler02} have used 204 SF galaxies
drawn from the Two Degree Field Galaxy Redshift Survey (2dFGRS) and
the 1.4~GHz NRAO VLA Sky Survey (NVSS) to derive their LF (see
\f{fig:lfcomp} ). However, to obtain the best fit parameters to
eq.~\ref{eq:sadler}, they combined this sample (that constrains well
the high luminosity end of the LF; see \f{fig:lfcomp} ) with the RSA
galaxy sample (in order to sample the low luminosity end of the LF).

The differences between the two local LFs are illustrated in
\f{fig:lfcomp} . 
There is a discrepancy between the two analytical representations of
  the radio LFs at both the high and low luminosity ends. However, the
  discrepancy affecting most severely the star formation rate density,
  that we aim to derive here, is the different position of the LF's
  turn-over point ($L_*$) given by the two analytic forms. This yields
  a difference of 10-50\% in the star formation rate density integral
  (see e.g.\ \f{fig:ld} \ and \f{fig:sfrdradio} ) as the luminosity
  range encompassing the turn-over point ($10^{19} \lesssim
  L_\mathrm{1.4GHz} \lesssim 10^{23}$~W~Hz$^{-1}$) contributes the
  most ($\sim95\%$) to the integral. It is important to note that the
  2dFGRS and NVSS data used by \citet{sadler02} sample more precisely
  the high-luminosity end (see \f{fig:lfcomp} ) of the local radio LF
  when compared to the RSA sample \citep{condon89}. Hence, in
  \s{sec:ulirgs} \ we will use the former one to constrain the
  evolution of our most intensely SF galaxies (i.e.\
  $L_{1.4}\gtrsim2.34\times 10^{23}$~\WH). As the {\em total} star
  formation rate density in each redshift range is derived by
  integrating under the evolved LF curve (multiplied with star
  formation rate; see below) down to the faintest 1.4~GHz
  luminosities, we will take this difference of the local 20~cm LFs
  into account in further analysis.

\subsection{ The evolution of star forming galaxies }
\label{sec:evol}

\begin{figure}[ht!]
{\center
\includegraphics[bb = 100 330 450 762, width=\columnwidth]{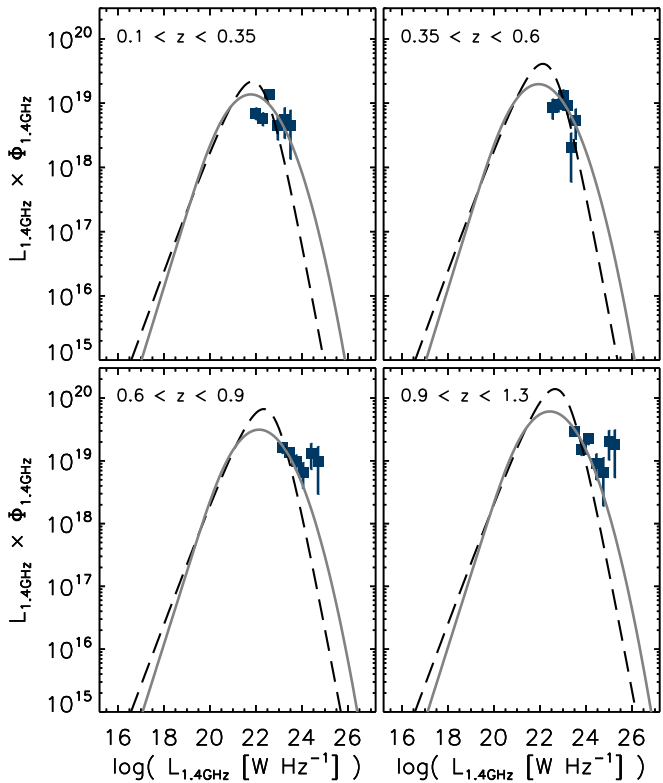}
\caption{ Luminosity density for VLA-COSMOS star forming galaxies
  (filled squares) in 4 redshift bins. The solid and dashed curves
  correspond to the best fit LFs in each redshift bin using the local
  LF given in \citet{sadler02} and \citet{condon89},
  respectively. Note that the first appears to describe the
  bright-luminosity end more consistently with our data.
  \label{fig:ld}}
}
\end{figure}

We parameterize the evolution of the VLA-COSMOS SF galaxy LF by pure
luminosity evolution:

\begin{equation}
\Phi_z(L) = \Phi_{z=0} \left[ \frac{L} {(1+z)^{\alpha_L}} \right]
\end{equation}
where $\alpha_L$ is the characteristic luminosity evolution parameter,
and $\Phi_z$ is the luminosity function at redshift $z$.  We derive
the evolution by summing the $\chi^2$ distributions obtained for a
large range of fixed $\alpha_L$ in each redshift bin (excluding our
first -- local -- redshift bin).  The uncertainty in $\alpha_L$ is
then taken to be the $2\sigma$ error obtained from the $\chi^2$
statistics. Our results yield a pure luminosity evolution with i)
$\alpha_L=2.1\pm0.2$, when the \citet{sadler02} local LF is used, and
ii) $\alpha_L=2.5\pm0.1$ when the \citet{condon89} local LF is
used. The different evolution parameters are a natural consequence of
the different slopes of the two local LFs in the luminosity range that
is constrained by the VLA-COSMOS data (see
Figs.~\ref{fig:lfcomp}~and~\ref{fig:ld}).

\citet{haarsma00} have found that a pure luminosity evolution with
$\alpha_L\approx2.74$ is a good representation of the evolution of
their radio-selected SF galaxies [no uncertainties were associated
with this estimate]. They used the \citet{condon89} local LF as the
basis for deriving their evolution. Given i) their much smaller sample
size, and ii) that their LF, when corrected for cosmology, agrees well
with the one derived here (see \f{fig:lf} ), these two results are in
good agreement.  Further, \citet{cowie04} find an evolution of
the SF galaxy LF (using the \citealt{condon89} local LF) consistent
with $\alpha_L$ of 3. The SF LFs, as well as the evolution, derived
here agree with their findings (c.f.\ Fig~3 in \citealt{cowie04}).
Our results are also consistent with the overall evolution of star
forming galaxies obtained by \citet[$\alpha_L = 2.7 \pm
0.6$]{hopkins04}, that has been shown to fit well the evolution of
radio selected star forming galaxies at low redshifts
\citep[$z\leq0.5$;][Phoenix Deep Survey]{afonso05}.

%
%

\section{ The cosmic star formation history (CSFH) }
\label{sec:csfh}

\subsection{The total cosmic star formation history}
\label{sec:csfhall}

As the star formation rate density (SFRD) is derived by integrating
the luminosity density, in \f{fig:ld} \ we show the luminosity density
for our 4 redshift bins. The shown curves are the two local 20~cm LFs
\citep{condon89, sadler02}, purely evolved in luminosity, and best fit
to the VLA-COSMOS data in each redshift range.  Prior to integration,
we convert the 1.4~GHz radio luminosity to star formation rates
($\psi$, in \Msolyr ), using the calibration given in \citet{bell03},
based on the total IR -- radio correlation:
\begin{equation} \label{eq:bell}
\mathrm{\psi \,\, [\mathrm{M}_{\odot}\, \mathrm{yr}^{-1}]} = 
\left\{ \begin{array}{lr}
      5.52\times 10^{-22}\,\, \mathrm{L_{1.4}}\,\, , & 
                                          \mathrm{L_{1.4}} > \mathrm{L_c} \\
      \frac{ 5.52\times 10^{-22} }   
           { 0.1 + 0.9 \left( 
                              \mathrm{L_{1.4}} / L_c \right)^{0.3}
           }
           \,\, \mathrm{L_{1.4}}\,\, , & 
                                         \mathrm{L_{1.4}}\le \mathrm{L_c} 
      \end{array} \right.
\end{equation}
where $\mathrm{L_c}=6.4\times10^{21}$~\WH, and $\mathrm{L_{1.4}}$
is the 1.4~GHz radio luminosity in units of \WH.  This calibration
uses a Salpeter initial mass function (IMF $\propto M^{-2.35}$) from
0.1 -- 100~\Msol . After the conversion, we compute the star formation
rate density (SFRD) for a given redshift bin as $\int
\psi\left(L\right)\, \Phi_z\left(L\right)\, dL$, where $\Phi_z$ is the
evolved radio LF best fit to the data in each redshift range (see
curves in \f{fig:ld} ).

In \f{fig:sfrdnum} \ we first show the SFRD, obtained by numerically
integrating the VLA-COSMOS data (open squares) within the sampled
luminosity range, with no attempt made to extrapolate towards the
faint or bright luminosity ends using the evolved local LF. This
eliminates any assumption, and yields robust lower limits purely
obtained from the data. The dependence of the derived SFRD on the
faint/bright end extrapolation is illustrated by the (solid and
dashed) curves in \f{fig:sfrdnum} , which were obtained using the
average pure luminosity evolution of the two local LFs derived in
\s{sec:evol} , and integrating over the entire SFRD curve. Hence, the
largest uncertainty in the derivation of the SFRD based on radio data
arises from the uncertainty in the shape of the local radio LF
(assumed not to change with redshift), and the associated
extrapolation below the faintest luminosity sampled by the data.

\begin{figure}[ht!]
\center{
\includegraphics[bb = 54 360 486 792, width=\columnwidth]{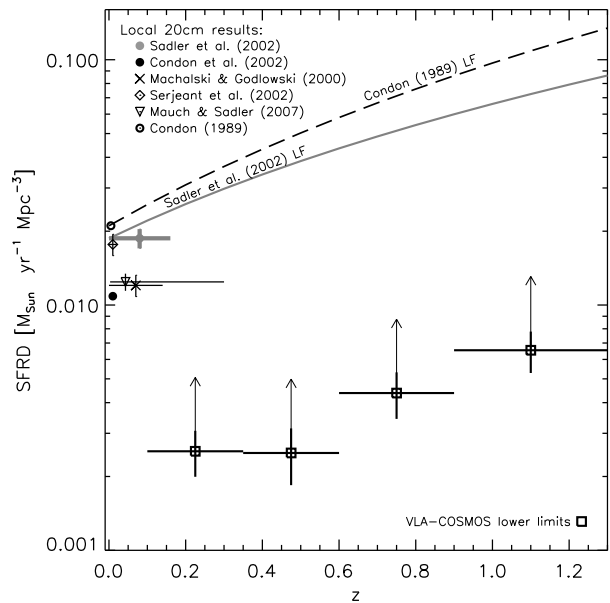}
\caption{ Star formation rate density (SFRD) as a function of
redshift. Open squares represent the SFRD derived from pure VLA-COSMOS
data, without any extrapolations towards the faint or bright
luminosity end. Therefore, they present strict lower limits, and are
indicated with arrows. The two curves shown correspond to the average
evolution of the VLA-COSMOS SF galaxies derived using two different
local radio LFs \citep{condon89,sadler02}. Local SFRDs derived by
various authors are also shown, and indicated in the panel.
  \label{fig:sfrdnum}}
}
\end{figure}

\begin{deluxetable}{c|c|c}
\tablecaption{ 1.4~GHz radio luminosity functions for VLA-COSMOS star
forming galaxies ($H_0=70,\, \Omega_M=0.3, \Omega_\Lambda = 0.7$).
\label{tab:LFs}
  }
\tablehead{
\colhead{ redshift} &
\colhead{$\mathrm{L_{1.4GHz}}$} &
\colhead{$\Phi $}\\
\colhead{range} &
\colhead{$[\mathrm{ W Hz^{-1}}]$} &
\colhead{$[\mathrm{Mpc^{-3} dex^{-1}]}$}
}
\startdata
                & $1.0\cdot10^{22}$ & $7.0\pm1.5\cdot10^{-4}$ \\
$0.1<z\leq0.35$ & $2.0\cdot10^{22}$ & $2.9\pm 0.7\cdot10^{-4}$ \\
                & $4.0\cdot10^{22}$ & $3.5\pm 0.6\cdot10^{-4}$ \\
                & $8.9\cdot10^{22}$ & $5.0\pm 2.1\cdot10^{-5}$ \\
                & $1.8\cdot10^{23}$ & $3.2\pm 1.6\cdot10^{-5}$ \\
                & $3.2\cdot10^{23}$ & $1.4\pm 1.0\cdot10^{-5}$ \\
\hline
                & $3.5\cdot10^{22}$ & $2.4\pm 0.9\cdot10^{-4}$ \\
$0.35<z\leq0.6$ & $6.3\cdot10^{22}$ & $1.5\pm 0.4\cdot10^{-4}$ \\
                & $1.0\cdot10^{23}$ & $1.3\pm 0.3\cdot10^{-4}$ \\
                & $1.6\cdot10^{23}$ & $5.8\pm 1.6\cdot10^{-5}$ \\
                & $2.2\cdot10^{23}$ & $9.0\pm 6.4\cdot10^{-6}$ \\
                & $3.5\cdot10^{23}$ & $1.5\pm 0.7\cdot10^{-6}$ \\
\hline
                & $1.4\cdot10^{23}$ & $1.2\pm 0.2\cdot10^{-5}$ \\
$0.6<z\leq0.9$  & $2.8\cdot10^{23}$ & $4.8\pm0.9\cdot10^{-5}$ \\
                & $5.6\cdot10^{23}$ & $1.7\pm0.5\cdot10^{-5}$\\
                & $1.1\cdot10^{24}$ & $5.8\pm2.6\cdot10^{-6}$\\
                & $2.5\cdot10^{24}$ & $5.2\pm2.3\cdot10^{-6}$\\
                & $5.0\cdot10^{24}$ & $2.0\pm1.4\cdot10^{-6}$\\
\hline
                & $3.2\cdot10^{23}$ & $9.2\pm1.5\cdot10^{-5}$\\
$0.9<z\leq1.3$  & $6.3\cdot10^{23}$ & $2.4\pm0.5\cdot10^{-5}$\\
                & $1.3\cdot10^{24}$ & $1.8\pm0.4\cdot10^{-5}$\\
                & $2.8\cdot10^{24}$ & $3.2\pm1.4\cdot10^{-6}$\\
                & $5.6\cdot10^{24}$ & $1.2\pm0.8\cdot10^{-6}$\\
                & $1.0\cdot10^{25}$ & $2.0\pm1.0\cdot10^{-6}$\\
                & $1.8\cdot10^{25}$ & $1.0\pm0.7\cdot10^{-6}$
\enddata
\end{deluxetable}

We compare our SFRD results (given in \t{tab:SFRDs} ), obtained by
integrating over the best fit evolved local LF in each redshift range
(see curves in \f{fig:ld} ), with other radio-based estimates in
\f{fig:sfrdradio} . The expected steep decline in the star formation
rate density since $z\sim1$ is reproduced by our data. Our results are
consistent within the errors with the results from \citet[who used the
\citealt{condon89} local LF]{haarsma00}, although their results are on
average higher. Note also that our statistical CSFH uncertainties
(bold red crosses in \f{fig:sfrdradio} ) are significantly smaller
compared to the \citet{haarsma00} results, as a result of the larger
sample utilized here.  Our results are also qualitatively consistent
with those obtained by \citet{ivison07}, based on radio image
stacking of UV selected galaxies in the AEGIS20 survey.

\begin{deluxetable}{c|c|c}
\tablecaption{ Star formation rate density (derived using the
  calibration given in \citealt{bell03}; see eq.~\ref{eq:bell}) from
  1.4~GHz VLA-COSMOS data ($H_0=70,\, \Omega_M=0.3, \Omega_\Lambda =
  0.7$).
\label{tab:SFRDs}
  }
\tablehead{
\colhead{redshift} & 
\colhead{\hspace{21mm} SFRD } &
\colhead{\hspace{-21mm}[\Msolyr ]}\\
\colhead{range} &
\colhead{\scriptsize{ \citet{condon89} LF }} &
\colhead{\scriptsize{ \citet{sadler02} LF }}
}
\startdata
$0.1<z\leq0.35$ & $0.025^{+0.001}_{-0.001}$ & $0.023^{+0.002}_{-0.002}$ \\
$0.35<z\leq0.6$ & $0.043^{+0.003}_{-0.003}$ & $0.032^{+0.003}_{-0.002}$ \\
$0.6<z\leq0.9$  & $0.067^{+0.003}_{-0.003}$ & $0.048^{+0.003}_{-0.004}$ \\
$0.9<z\leq1.3$  & $0.134^{+0.010}_{-0.009}$ & $0.088^{+0.005}_{-0.005}$
\enddata
\end{deluxetable}

\citet{seymour08} have used VLA/MERLIN radio frequency observations of
the 13$^\mathrm{H}$ XMM-Newton/Chandra Deep field to derive the cosmic
star formation history. Their findings are qualitatively consistent
with those from \citet[shown in \f{fig:sfrdradio} ]{haarsma00} when
they use the \citet{mauch07} local LF evolved in luminosity with an
a-priori set value of $\alpha_L=2.5$ (note that this local LF has a
lower normalization than \citealt{sadler02}; see the local results in
\f{fig:sfrdnum} \ and \f{fig:sfrdradio} ). Further, they used a
different radio luminosity to SFR relation that is consistent with
0.84 times the \citet{bell03} calibration used here (see
\citealt{seymour08} for details). Hence, this implies that their
results are significantly higher than the VLA-COSMOS results. Further,
given the differences between the local radio LFs outlined in
\s{sec:locLFs} , if \citet{seymour08} had chosen to use the
\citet{condon89} local LF with otherwise the same assumptions, they
would have obtained even higher SFRD values. We believe that the main
reason for the differences between our results and those by
\citet{seymour08} is likely a combination of i) their a-priori assumed
evolution of the local LF, contrary to constraining the evolution by
their data and ii) the inclusion of a fraction of lower power (i.e.\
radio quiet) AGN in their star forming galaxy sample, while our sample
may over-subtract composite objects (see also \s{sec:discussion} ).

In \f{fig:sfrdall} \ we compare the VLA-COSMOS derived CSFH data with
results from previous studies based on a range of SF estimators -- UV,
optical, FIR, total IR, and radio. A luminosity-dependent obscuration
correction was used where necessary (see \citealt[and references
therein]{hopkins04}).  Overall, our derived CSFH agrees with the
general trend of a rapid decline by almost an order of magnitude in
the cosmic star formation rate density since $z\sim1$. A possibly
slower decline is suggested by our data if the \citet{sadler02} LF is
used. However, as the uncertainties due to the faint-end
extrapolation are significant, no robust conclusions can be made at
this point.

\begin{figure}[ht!]
\center{
\includegraphics[bb = 54 360 486 792, width=\columnwidth]{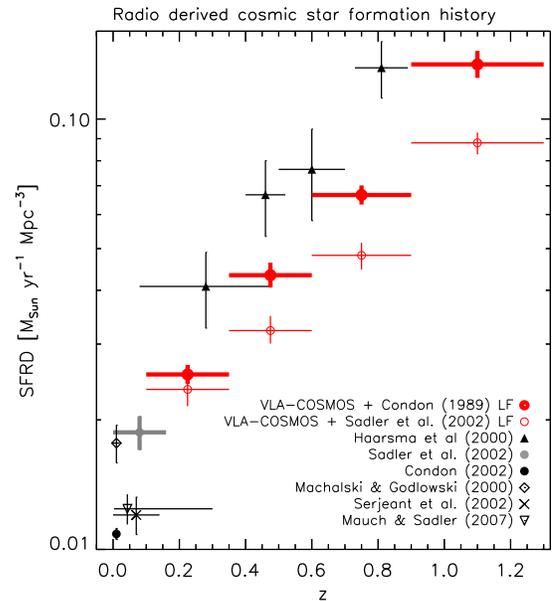}
\caption{ Comparison of radio derived cosmic star formation
histories. The different samples are indicated in the panel. The bold
red circles show the VLA-COSMOS results, when the \citet{condon89}
local LF is used; open red circles show the results when the
\citet{sadler02} local LF is used. Results based on other radio-based
studies are also shown, and indicated in the panel.
  \label{fig:sfrdradio}}  }
\vspace{1.5mm}
\end{figure}

\subsection{ The CSFH of intensely star forming galaxies }
\label{sec:ulirgs}

The VLA-COSMOS SF galaxy sample constrains well the high end of the LF for SF
galaxies.  Given the 2\sqdegs\ VLA-COSMOS field the comoving volume sampled up
to $z=1.3$ is $\sim1.3\times10^7$~Mpc$^3$, corresponding roughly to the volume
observed locally by SDSS (DR1).  Thus, for the first time this allows a robust
derivation of the CSFH for galaxies forming stars at rates of
$\gtrsim100$~\Msolyr\ out to $z=1.3$. Such radio selected galaxies are
equivalents to the ultra-luminous IR galaxies (ULIRGs,
$\mathrm{L_{IR}}>10^{12}$~\lsun), and it is noteworthy that the VLA-COSMOS
survey is sensitive to a {\em complete} sample of these galaxies out to
$z\sim1$ (see Fig.~16 in S08).

In order to derive the evolution of the SFRD at the high-luminosity
end, we integrate the SFRD curve, obtained from the best fit pure
radio luminosity evolution in each redshift bin (see \f{fig:ld} ),
only for our SF galaxies that have $L_{1.4}\gtrsim2.34\times
10^{23}$~\WH, which corresponds to $\mathrm{L_{IR}}>10^{12}$~\lsun\
given the adopted total IR -- radio correlation \citep{bell03}. For
this we use the local LF given by \citet{sadler02} as it appears to be
better suited for the high-luminosity end compared to the
\citet{condon89} LF (see \f{fig:lfcomp} ). Note that for these highly
luminous galaxies the extrapolation uncertainties are not as
significant as for the overall SFRD, as this sample is almost complete
in all three high redshift ranges. A small extrapolation to the faint
end, given the form of the evolved local LF, is necessary only in the
last redshift bin. For a consistent comparison between our radio and
IR (LF05) results we convert the total IR luminosity to star formation
rates consistently using the calibration given by \citet{bell03}.  The
results are shown in \f{fig:sfrdulirg} .
%
%
The evolution of our {\em star forming} ULIRGs is consistent with the
lower envelope derived by LF05. However, it is marginally flatter,
suggesting a slower evolution of star forming ULIRG galaxies since
$z=1.3$. This will be further discussed in the next section.

\begin{figure}
\includegraphics[bb = 54 360 486 790, width=\columnwidth]
                                           {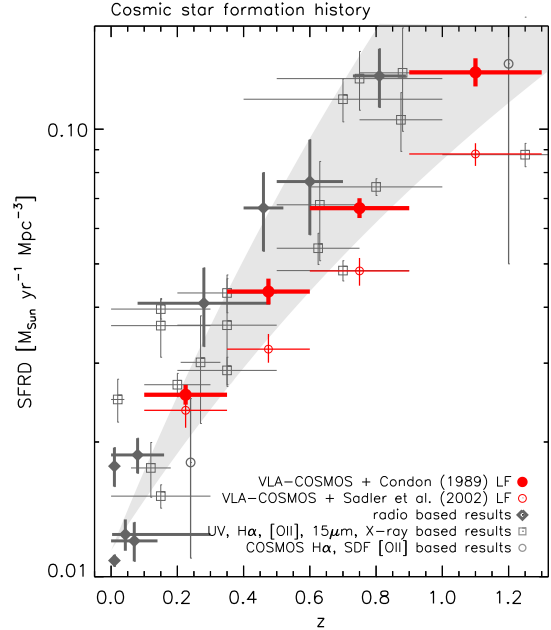}
\caption{ Comparison of the VLA-COSMOS derived cosmic star formation
history (red symbols; see \f{fig:sfrdradio} \ for details) with
other-wavelength based results. The compilation of CSFHs contains UV-,
H$\alpha-$, FIR-, and X-ray- based results, corrected for
dust-obscuration using luminosity-dependent corrections (open gray
squares; see \citealt{hopkins04} for details). Thick gray diamonds
denote local radio estimates
(\citealt{machalski00,condon02,sadler02,serjeant02}).  For all radio
data the 1.4~GHz luminosity to star formation rate calibration given
in \citet[see eq.~\ref{eq:bell}]{bell03} is used. The gray-shaded area
shows the cosmic star formation history derived by LF05, based on
24~\mic\ data. The open gray circles denote i) the H$\alpha$ derived
SFRD at $z=0.24$ in the COSMOS field, corrected for both
dust-obscuration and AGN contribution \citep{shioya08}, and ii) the
dust-obscuration corrected SFRD at $z=1.2$ derived using [OII]
emission galaxies in the Subaru Deep Field (SDF; see for
\citealt{takahashi07} details).
  \label{fig:sfrdall}}\vspace{2mm}
\end{figure}

\begin{figure}[h]
\includegraphics[bb = 54 370 486 770, width=\columnwidth]{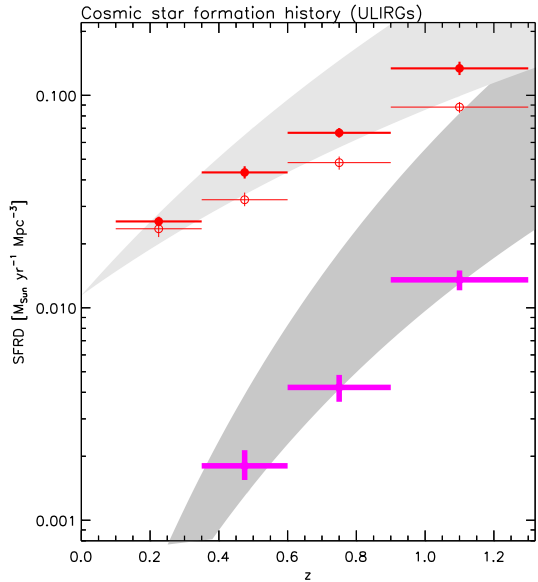}
\caption{ Cosmic star formation history (CSFH) derived from VLA-COSMOS
  data for the overall population (red symbols as in
  \f{fig:sfrdall} ), and only for star forming ULIRGs (magenta
  symbols; derived using the \citealt{sadler02} local LF evolved in
  luminosity in each redshift bin).  Also shown is the CSFH for the
  entire (light-gray shaded curve), and the ULIRG population
  (dark-gray shaded curve), derived from the evolved total IR
  luminosity function (LF05). The IR- and radio- based star formation
  rates have been put on the same relative scale (see text for
  details). Note that the VLA-COSMOS data suggest a slightly slower
  evolution of the {\em star forming} ULIRG population with redshift.
  \label{fig:sfrdulirg}}
\end{figure}

\vspace{0.5cm}
\section{ Discussion }
\label{sec:discussion}

\subsection{The evolution of the most intensely star forming galaxies}

Making use of our large statistical sample of radio-selected {\em star
  forming} ULIRGs complete out to $z\sim1$ we have derived the CSFH of
  the most intensely star forming galaxies
  ($\gtrsim100$~\Msol~yr$^{-1}$) out to $z=1.3$. Our evolution of the
  cosmic star formation rate in star forming ULIRGs qualitatively
  agrees with previous MIR-based results (LF05; see \f{fig:sfrdulirg}
  ). However, we find a slightly slower evolution than predicted by
  the MIR results. The major cause for this difference is currently
  unclear, nonetheless there are likely three effects that may
  contribute:


  i) No attempt has been made to minimize the AGN contamination in
  the 24~\mic - selected sample possibly causing an overestimate in
  the MIR-derived SFRD evolution for ULIRGs. For example,
  \citet{caputi07} have found $>10\%$ of 24~\mic-AGN at $z\sim1$, and
  a factor of 2 more at $z\sim2$, suggesting that the AGN fraction in
  MIR selected samples increases with redshift. The AGN fraction in MIR
  samples may also be a function of stellar mass. Although at higher
  redshift than analyzed here (implying a different cosmological era),
  \citet{daddi07} have demonstrated that at $z\sim2$ the MIR AGN
  fraction is indeed a function of stellar mass, and reaches $\sim50-60\%$
  for masses $>4\times10^{10}$~\Msol . The median stellar mass of the
  VLA-COSMOS star forming galaxies (obtained via SED fitting; see S08
  for details) is $\sim7\times10^{10}$~\Msol . 

  ii) Particular care was taken to separate the VLA-COSMOS population
  into SF and AGN galaxies. Nonetheless, it has to be noted that some
  uncertainty, due to AGN contamination as well as incompleteness of
  the star forming galaxy sample exists, especially on the
  galaxy-by-galaxy and composite (SF plus AGN) galaxy level (see \citealt{smo08} for
  details). It is also worth noting that two significant large scale
  structure components exist in our highest redshift bin
  (\citealt{scoville07a}, Scoville et al., in prep) that may affect the
  fraction of SF and AGN galaxies present in this particular redshift
  range.

  iii) The local IRAS total IR and the Sadler et al.\ radio LFs,
  used for the derivation of these results, are not perfectly
  similar. There may also be a volume density excess of SF galaxies
  with high radio luminosities at $z>0.6$, compared to the Sadler et
  al.\ LF evolved only in luminosity. In addition, the total IR --
  radio correlation, used to select ULIRGs from our radio sample,
  carries its own uncertainty and intrinsic astrophysical scatter
  \citep{bressan02, bell03}.  Therefore, it is not immediately obvious
  whether the same results would be expected based on both -- radio
  and MIR -- star formation indicators. 

  In summary, it is encouraging that the same qualitative behavior of
  the evolution of the ULIRG population is observed with the two
  independent, radio and MIR, SFR indicators. However, further
  dedicated studies
  of the details will prove most interesting in understanding the
  quantitative differences seen in \f{fig:sfrdulirg} .

\subsection{The rapid decline of the CSFH since $z\sim1$}

Our overall CSFH (\f{fig:sfrdall} ) agrees well with past
findings, when these are corrected for dust-obscuration as
needed. This verifies the assumptions about large dust-obscuration
corrections required, especially for short-wavelength (e.g.\ UV) star
formation tracers. Our radio data independently confirm the $\sim1$
order of magnitude decline in the cosmic star formation rate since
$z\sim1$.

Based on UV and IR based SFR/morphology studies \citep{wolf05, bell05,
melbourne05, hammer05, zamojski07} this rapid decline in the overall
cosmic star formation history is expected to be driven by normal
spiral galaxies.  For example, \citet{bell05} have performed a detailed
morphological study of a galaxy sample ($z=0.7$) with SFRs
$\gtrsim10$~\Msolyr . They have demonstrated 
that physical processes that do not substantially affect galaxy
morphology, such as minor mergers, gas consumption and weak
interactions with satellite galaxies, may be most important for the
rapid decline in the overall CSFH \citep[see also e.g.\
][]{somerville01}. 

The sample analyzed here, however, is most sensitive to ULIRGs
(SFR~$\gtrsim100$~\Msolyr ), and those are the systems a priori
expected to be a reflection of galaxy merging \citep[based on local
ULIRG morphology studies; ][]{sanders96}. This implies that the rapid
decline in CSFH of our ULIRGs (\f{fig:sfrdulirg} ) may be more
affected by galaxy mergers than the overall CSFH decline since
$z\sim1$. Intriguingly, our derived LF evolution is very similar to
the evolution of the galaxy close-pair fraction, derived for bright
galaxies \citep[$>L_V^*$; ][]{kartaltepe07} in the COSMOS field, well
described with a power law with an index of $3.1\pm0.1$ (or
$2.2\pm0.1$ when pure luminosity evolution is considered; see also
e.g.\ \citealt{lotz08}). This strong evolution of the bright
close-pair fraction, combined with a similarly strong evolution of the
CSFH of ULIRGs derived here suggests that major mergers may play an
important role in the rapid decline of the CSFH since $z\sim1$ for the
most intensely star forming galaxies. Thus, the observed steep
evolution of our ULIRG population may be a good record of merger rate
evolution, combined with gas content evolution.

\section{Summary}

We have derived the cosmic star formation history out to $z=1.3$ using
to date the largest sample of {\em radio-selected} star forming
galaxies observed at 1.4~GHz (20~cm) in the VLA-COSMOS survey. The
large increase in the number of radio selected SF galaxies out to high
redshift, compared to previous studies, allowed us to constrain well the
evolution of the 1.4~GHz luminosity function for radio-selected {\em
  star forming} galaxies, as well as to  reduce significantly the
statistical uncertainties of the radio-derived CSFH. We find that the
uncertainties are ruled by the differences in the shape of the local
radio LFs present in the literature. A pure radio luminosity
evolution of VLA-COSMOS star forming galaxies is well described with
$L_*\propto(1+z)^{2.1\pm0.2}$, when evolving the \citet{sadler02}
local LF, or with $L_*\propto(1+z)^{2.5\pm0.1}$ when evolving the
\citet{condon89} local LF. Although encompassing a
relatively broad range, both values are consistent with previously
derived evolution of star forming galaxies (e.g.\ \citealt{condon02,
  hopkins04, cowie04, afonso05}).

Our overall CSFH agrees well with past findings, when these are
corrected for dust-obscuration where needed. This verifies the
assumptions about large dust-obscuration corrections required,
especially for short-wavelength (e.g.\ UV) star formation tracers.
Making use of our large statistical sample of radio-selected {\em star
forming} ULIRGs complete out to $z\sim1$ we have robustly constrained
the high-end of the SF galaxy LF at different cosmic times. Using
these we have derived the CSFH of the most intensely star forming
galaxies ($\gtrsim100$~\Msol~yr$^{-1}$; i.e.\ {\em star forming}
ULIRGs) out to $z=1.3$. We find an, on average, slower evolution of
the cosmic star formation rate in star forming ULIRGs than predicted
by MIR results consistent with the fraction of star forming galaxies
in MIR samples likely becoming lower with increasing redshift and/or
stellar mass.

\acknowledgments CC acknowledges support from the Max-Planck Society
and the Alexander von Humboldt Foundation through the
Max-Planck-Forschungspreis 2005. GZ acknowledges support from two
contracts PRIN-INAF 2005 and 2007. TP acknowledges support from PSC-CUNY
grant \# 69612-00 38.

{}

\end{document}